# PRESENT STATUS OF SOURCE DEVELOPMENT STATION AT UVSOR-III *

Najmeh Sadat Mirian[#], Jun-ichiro Yamazaki, Kenji Hayashi, Masahiro Katoh, UVSOR, Okazaki, Japan
Masahito Hosaka, Yoshifumi Takashima, Nagoya University, Nagoya, Japan
Naoto Yamamoto, Taro Konomi, KEK, Tsukuba, Japan
Heishun Zen, Kyoto University, Kyoto, Japan

## Abstract

Construction and development of a source development station are in progress at UVSOR-III, a 750 MeV electron storage ring. It is equipped with an optical klystron type undulator system, a mode lock Ti:Sa Laser system, a dedicated beam-line for visible-VUV radiation and a parasitic beam-line for THz radiation. New light port to extract edge radiation was constructed recently. An optical cavity for a resonator free electron laser is currently being reconstructed. Some experiments such as coherent THz radiation, coherent harmonic radiation, laser Compton Scattering gamma-rays and optical vortices are in progress.

## INTRODUCTION

UVSOR is a synchrotron light source, which was constructed in 1980's. Using a part of the ring, various light source technologies, such as resonator free electron laser [1] and its applications [2], coherent harmonic generation [3] and coherent synchrotron radiation via laser modulation [4], laser Compton scattering [5] have been developed. These research works had been carried out by parasitically using an undulator and a beam-line for photo-electron spectroscopy [6]. Under Quantum Beam Technology Program of MEXT in Japan, we started constructing a new experiment station dedicated for light source developments. FY2010, we created a new straight section by moving the injection line. FY2011, a new optical klystron was constructed and installed. FY2009-2010, the seed laser system was upgraded and moved to the new station. FY2011, two beam-lines dedicated for coherent light source development were constructed. In FY2012, another upgrade program for the storage ring was funded, in which all the bending magnets were replaced [7]. After this major upgrade, we started to call the machine UVSOR-III. Because we had to pay a lot of efforts for the machine conditioning, we have to slow-down the construction of the source development station for a few years. In FY2014, the mirror chambers of the optical cavity were installed. The experiments have started on coherent THz edge radiation, optical vortex beam, and laser Compton scattering gamma-rays. In this paper, we will report the most recent status of the source development station at UVSOR-III.

___________________________________________
*Some parts of this work were supported by JSPS KAKENHI Grant Number 26286081, 26390111 and MEXT Quantum Beam Technology Program
#nsmirian@ims.ac.jp

## FACILITY STATUS

### Accelerators

The recent view of UVSOR-III storage ring is shown in FIG. 1. The main parameters of the ring are listed in Table 1. The ring is normally operated at 750 MeV for synchrotron radiation users in multi-bunch mode. On the other hand, in many of the source development studies, the ring is operated at lower energy (600~500MeV) and in single bunch mode. The studies are carried out in dedicated beam times for machine studies. Usually every weekend and Monday can be used for machine studies. In addition, a few weeks a year are usually reserved for machine studies.

The electron beam is supplied by an injector which consists of a 15 MeV linear accelerator and a full energy booster synchrotron. Top-up operation is possible, even for the low energy single bunch operation.

Since the major upgrade in 2012, we have observed that the threshold current of the transverse single bunch instability was lowered. Currently we can accumulate around 50 mA in a single bunch, however, it is difficult to accumulate more. This problem is currently under investigation.

The source development station was constructed by utilizing one of 4m straight section in the ring. It is comprised of an optical klystron, an optical cavity, a seed laser system and beam-lines. The layout of the accelerator part is shown in FIG. 2.

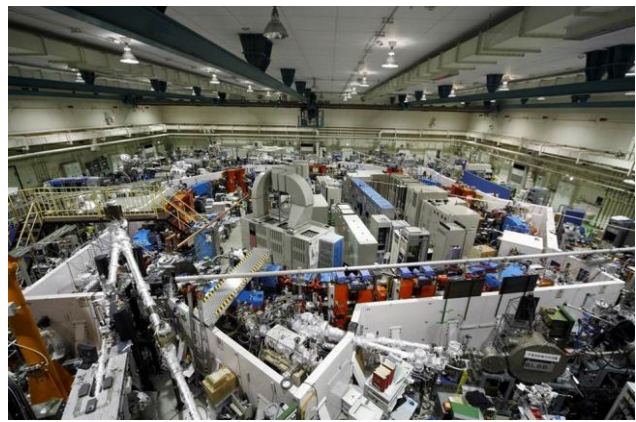

Figure 1: Recent View of UVSOR-III Storage Ring and Synchrotron Radiation beam-lines.

Table 1: Main Parameters of UVSOR-III

| | |
|---|---|
| Electron Energy | 750 MeV (max.) |
| Circumference | 53.2 m |
| Beam Current | 300 mA (multi-bunch) |
| | 50 mA (single bunch) |
| Emittance | 17.5 nm-rad |
| Energy Spread | $5.3 \times 10^{-4}$ |
| Betatron Tunes (x,y) | (3.75, 3.20) |
| Harmonic Number | 16 |
| RF Frequency | 90.1 MHz |
| RF Voltage | 120 kV |
| Momentum Compaction | 0.030 |
| Natural Bunch Length | 128 ps |

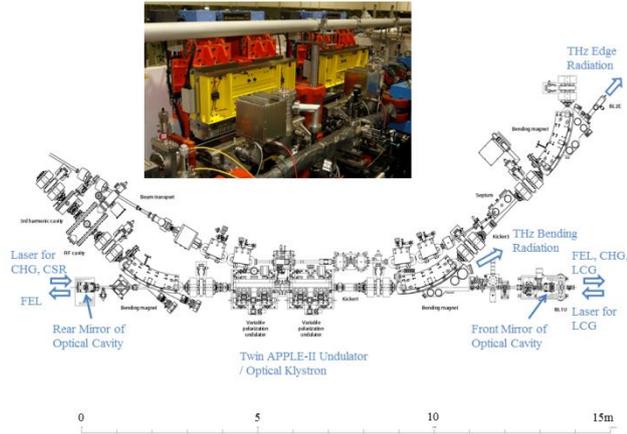

Figure 2: Layout of Source Development Station (Accelerator Part) with a picture of Optical Klystron. A part of the storage ring is shown. The electron beam is circulating counter clockwise (from left to right in this figure).

*Optical Klystron*

The optical klystron consists of two identical variable polarization undulators of APPLE-II configuration and a buncher magnet, which is a three pole electromagnetic wiggler, in between. The magnetic gaps and phases of the two undulators as well as the field strength of the buncher can be changed independently. The undulators were designed so that the fundamental wavelength of the undulators in the planer mode can be tuned to 800 nm for the electron energy of 600 MeV and to 400 nm for 750MeV. The former wavelength is that of the Ti:Sa laser described in the next section and the latter its second harmonics. So far, the laser seeding is carried out at the fundamental wavelength. In future, the second harmonics will be used to carry out the seeding experiments during normal users beam times. The main parameters of the optical klystron are listed in Table 2.

Table 2: Main Parameters of Optical klystron

| | |
|---|---|
| Magnetic Configuration | APPLE-II |
| Period Length | 88 mm |
| Number of Periods | 10 + 10 |
| Max. R56 of Buncher | 67μm (600MeV) |
| Max. Deflection Parameter | 7.36 (horizontal) |
| | 4.93 (vertical) |
| | 4.06 (circular) |

*Laser*

The original laser system for the seeding experiment at UVSOR consisted of a Ti:Sa oscillator and a regenerative amplifier (COHERENT: legend HE), which was synchronized with the RF acceleration of the storage ring [4]. FY 2010, a multipath amplifier (COHERENT: Hidra-50) and a single-path amplifier (COHERENT: Legend-cryo) were added toward a higher pulse energy. The main parameters of the laser system are listed in Table 3.

During the construction of the new experiment station, the laser system was moved to a new site which is close to the downstream end of the undulator beam-line. For the seeding experiment, we have constructed a laser transport line which quides the laser beam to the upstream end of the undulator and, then, into the storage ring [8]. In this new configuration, we have observed laser beam position instability which is likely caused by mechanical vibrations of the optical components. It is observed that this causes significant instabilities of the coherent radiation intensities. Therefore we are currently testing a beam stabilization system based on a commercial product.

Table 2: Main Parameters of Laser System

| | | |
|---|---|---|
| Legend-HE | Pulse Energy | 2.5 mJ |
| | Pulse Width | 100fs-2ps |
| | Repetition Rate | 1 kHz |
| Legend-Cryo | Pulse Energy | 10 mJ |
| | Pulse Width | 100fs-2ps |
| | Repetition Rate | 1kHz |
| Hydra-50 | Pulse Energy | 50 mJ |
| | Pulse Width | 100fs-2ps |
| | Repetition Rate | 10Hz |

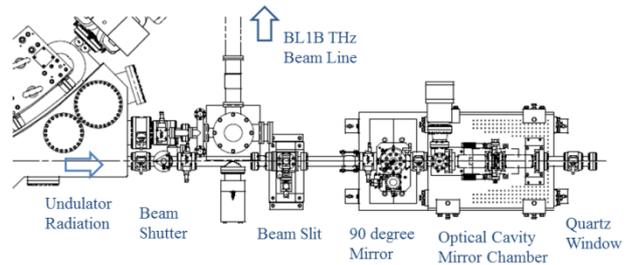

Figure 3: Beam Line Frontend at BL1U

*Beam Lines*

In spring 2015, the front end of the beam line BL1U, which is dedicated for extracting the FEL or CHG radiation or LCS gamma-rays and injecting laser beams for LCS experiments, was improved as shown in Figure 3. Just after the light port on the bending magnet chamber, there are a manual gate valve, a beam shutter and a pneumatic gate valve as in the other synchrotron radiation beam lines at UVSOR-III. Just downstream of these, there is a beam slit, which is to cut the unnecessary part of the radiation. This slit was prepared for future utilization of the coherent radiation. After the slit, a dismountable 90 degree mirror was installed, which is to extract the light beams and transport those to an optical bench for diagnostics and to lead the laser beam for the LCS experiment to the storage ring. After the mirror, a mirror chamber of the optical resonator for the FEL was installed. The exit of the mirror chamber was sealed with a quartz window. Currently the mirror for the optical cavity is not mounted. Therefore, the undulator radiation can be extracted directly through the window without using mirrors. This may be useful to precisely investigate the phase properties of the undulator radiation.

Another dedicated beam line BL2E was also constructed in FY2014. This beam line is located at the second bending magnet from the undulator, as shown in Figure 2. This beam-line utilizes a small light port on the zero degree line of the bending magnet. A water cooled Cu mirror reflects the edge radiation from the bending magnet upwards. The radiation was extracted to the air through a quartz window. In adding to this beam line, there is another terahertz beam line BL1B, which is normally used by terahertz synchrotron radiation users, but can be used for the coherent terahertz experiment. In this beam line, a magic mirror is installed which collects infrared and terahertz synchrotron radiation from the bending magnet with a very wide aperture, 244x80 mrad$^2$. A Martin-Puplett type interferometer (JASCO FARIS-1) is also equipped.

## SUMMARY AND PROSPECTS

A light source development station was constructed and is being developed at the UVSOR-III electron storage ring. Currently, coherent terahertz radiation experiment based on laser modulation technique, laser Compton scattering experiment and optical vortices beam study using the APPLE-II undulators in tandem are in progress. Resonator free electron laser experiment will be re-started in near future toward intra-cavity laser Compton scattering experiment. The combined use of multi-photon beams such as terahertz-pump and VUV-CHG probe experiments is under preparation.

## ACKNOWLEDGMENT

The authors wish to thank all the collaborators in the source development studies, from Lille U., Synchrotron Soleil, Dortmund U., Hiroshima U., Osaka U., Yokohama N. U. and KEK.